# RECENT DEVELOPMENTS IN MEMS-BASED MICRO FUEL CELLS

*Tristan Pichonat[1], Bernard Gauthier-Manuel*

Institut FEMTO-ST, Département LPMO, CNRS UMR 6174,
32 avenue de l'observatoire, 25044 Besançon Cedex, France

**ABSTRACT**

Micro fuel cells (µ-FC) represent promising power sources for portable applications. Today, one of the technological ways to make µ-FC is to have recourse to standard microfabrication techniques used in the fabrication of micro electromechanical systems (MEMS). This paper shows an overview on the applications of MEMS techniques on miniature FC by presenting several solutions developed throughout the world. It also describes the latest developments of a new porous silicon-based miniature fuel cell. Using a silane grafted on an inorganic porous media as the proton-exchange membrane instead of a common ionomer such as Nafion®, the fuel cell achieved a maximum power density of 58 mW $cm^{-2}$ at room temperature with hydrogen as fuel.

## 1. INTRODUCTION

Today, the development and miniaturization of portable devices such as cellular phones, PDA or any mobile electronics involve new needs in terms of energy supply and autonomy. In the context of these applications, miniature fuel cells (FC) represent a true alternative to common batteries, e.g. Li-ion batteries, due to their large energy density, their theoretical simplicity (easy and immediate recharge by refilling with the fuel) and their inherent non-polluting aspect.

For the range of powers required (<10 W), recent work [1-3] has shown a growing interest for the development of MEMS and CMOS compatible processes for small FC fabrication with silicon, stainless steel foils or polymer technologies. Beyond standard silicon, the use of porous silicon (PS) has also been demonstrated as an attractive material for gas diffusion layer [4], support for catalyst [5-6] or as a reformer [7] in the case of FC. Recent papers [8-9] have also shown the relevance of using PS as the membrane of a FC.

In this paper, we will focus on MEMS-based FC and try to show some of the most recent developments and technologies applied to this kind of miniature FC. We will also present the latest developments concerning a novel silane-grafted porous silicon-based technology for small FC conducted at the FEMTO-ST Institute.

## 2. MEMS-BASED FUEL CELLS

Since the 1980's, the MEMS technologies have been fully developed at the various requests of microsensors, micro actuators, optical and biomedical systems, and microfluidics. In the future prospects of miniaturization and mass production of small FC, the use of microfabrication techniques and materials appears naturally to be one of the promising solutions to improve the performances of the FC, reducing their fabrication costs and enabling their easier integration with other electronic devices.

In the domain of MEMS-based miniature FC, various solutions have been reported. Since the performances do not seem to be the better criteria to classify FC – actually we can notice in the literature that the power densities of the miniature FC range from a few tenths of µW $cm^{-2}$ up to several hundreds of mW $cm^{-2}$ – and as the structures are often the same – two micromachined plates for fuel delivery and current collectors sandwiching a membrane electrodes assembly (MEA) – we will index the different FC by similar materials. Nevertheless we will linger on miniature FC with the best performances. In this paper, we will not focus on the different fuels used (hydrogen for Proton-Exchange-Membrane FC or PEMFC, methanol for Direct Methanol FC or DMFC, ethanol, formic acid) but we will precise for each FC described the fuel supply conditions.

### 2.1. The silicon way

As the base material for MEMS technologies, Si is also the most common material encountered in MEMS-based FC. Its properties and the microfabrication techniques associated to it such as photolithography, wet or dry etching, depositing (sputtering, CVD, thermal oxidation, etc.) are now well-known and mastered. Another advantage of Si-based FC may also be to facilitate the possible integration of the FC with other electronic devices on the same chip.

---

[1] T.P. is now with the Silicon Microsystems Group, IEMN, Avenue H. Poincaré, BP 60069, F-59652 Villeneuve d'Ascq, FRANCE.
E-mail: tristan.pichonat@isen.iemn.univ-lille1.fr

Since 2000, Meyers *et al* (Lucent Technologies) have proposed two alternative designs using Si – a classical bipolar using separate Si wafers for the cathode and the anode and a less effective monolithic design that integrated the two electrodes onto the same Si surface [4,10]. In the bipolar design, both electrodes were constructed from conductive Si wafers. The reactants were distributed through a series of tunnels created by first forming a PS layer and then electropolishing away the Si beneath the porous film. The FC was completed by adding a catalyst film on top of the tunnels and finally by casting a Nafion® solution. Two of these membrane-electrode structures were made and then sandwiched together. A power density of 60 mW cm$^{-2}$ was announced for the bipolar design with H$_2$ supply.

Unlike Meyer and Maynard, Lee *et al* (Stanford Univ.) proposed a 'flip-flop' µ-FC design where both electrodes are present on the same face [2]. If this design does provide ease of manufacturing by allowing in-plane electrical connectivity, it complicates the gas management. Instead of electrons being routed from front to back, gasses must be routed in crossing patterns, significantly complicating the fabrication process and sealing. Peak power in a four-cell assembly achieved was still 40 mW cm$^{-2}$ with H$_2$ as fuel.

Min *et al* (Tohoky Univ, Japan) reported a variant of this design proposing two structures of µ-PEFC using microfabrication techniques [11], the "alternatively inverted structure" and the "coplanar structure". These structures use Si substrates with porous SiO$_2$ layers with Pt-based catalytic electrodes and gas feed holes, glass substrates with micro-gas channels, and a polymer membrane (Flemion® S). In spite of a reported enhancement [12], the FC reached poor results (only 0.8 mW cm$^{-2}$ for the alternatively inverted structure).

Yu *et al* (Hong Kong Univ.) described a miniature FC consisting in a MEA between two micromachined Si substrates [13]. By sandwiching Cu between layers of gold, they were able to decrease the internal resistance of the thin-film current collectors, which involved an increase performance of the FC, achieving a peak power density of 193 mW cm$^{-2}$ with H$_2$ and O$_2$.

Yen *et al* [14] (Univ. of California / Pennsylvania State Univ.) presented a bipolar Si-based micro DMFC with a MEA (Nafion®-112 membrane) integrated and micro channels 750 µm wide and 400 µm deep fabricated using Si micromachining (DRIE). This µDMFC with an active area of 1.625 cm$^2$ has been characterized at near room temperature (RT), showing a maximum power density of 47.2 mW cm$^{-2}$ at 60°C when 1M methanol was fed but only 14.3 mW cm$^{-2}$ at RT. Since then, Lu *et al* [15] have enhanced the performance of the µDMFC to a maximum power density of 16 mW cm$^{-2}$ at RT and 50 mW cm$^{-2}$ at 60°C with 2M and 4M methanol supply with a modified anode backing structure enabling to reduce methanol crossover. We will see the last evolution of their technology in the paragraph (2.2).

Yeom *et al* [16] (Univ. of Illinois) reported the fabrication of a monolithic Si-based microscale MEA consisting of two Si electrodes, with catalyst deposited directly on them, supporting a Nafion®-112 membrane in-between. The electrodes are identically gold-covered for current collecting, and are covered with electrodeposited Pt black. The electrodes and the Nafion® membrane are sandwiched and hot-pressed to form the MEA. The complete fuel cells have been tested with various fuels: H$_2$, methanol and formic acid and reached 35 mW cm$^{-2}$, 0.38 mW cm$^{-2}$ and 17 mW cm$^{-2}$ respectively at RT with forced O$_2$. More recently, performances with formic acid as fuel were increased up to 28 mW cm$^{-2}$ with electrodeposited Pd-containing catalyst at the anode[17].

Aravamudhan *et al* [18] (Univ. of South Florida) presented a FC powered by ethanol at RT. The electrodes have been fabricated using macro-porous Si technology. The pores developed act both as micro-capillaries/wicking structures and as built-in fuel reservoir, reducing the size of the FC. The pore sizes dictate the pumping/priming pressure in the FC. The PS electrode thus eliminates the need for an active external fuel pump. The structure of the MEA consists of two PS electrodes sandwiching a Nafion-115™ membrane. Pt was deposited on both the electrodes micro-columns to act both as an electrocatalyst and as a current collector. The FC reached a maximum power density of 8.1 mW cm$^{-2}$ by supplying 8.5 M ethanol solution at RT.

Recently, Yao *et al* [19] (Carnegie Mellon Univ.) were currently working on a RT DMFC to produce a net output of 10 mW for continuous power generation. Their works focus on the design of the complete system including water management at the cathode, micro pumps and valves, CO$_2$ gas separation and other fluidic devices. A passive gas bubble separator removes CO$_2$ from the methanol chamber at the anode side. The back planes of both electrodes are made of Si wafers with an array of etched micro-sized holes. Nano-tubes catalysts are fabricated on the planes. A 3 % methanol solution at the anode and the air at the cathode are driven by natural convection instead of being pumped. A micro pump sends water back to the anode side. With 25 mW cm$^{-2}$, the total MEA area around 1 cm² can provide enough power to a 10 mW microsensor along with the extra power needed for internal use, such as water pumping, electronic controls and process conditioning.

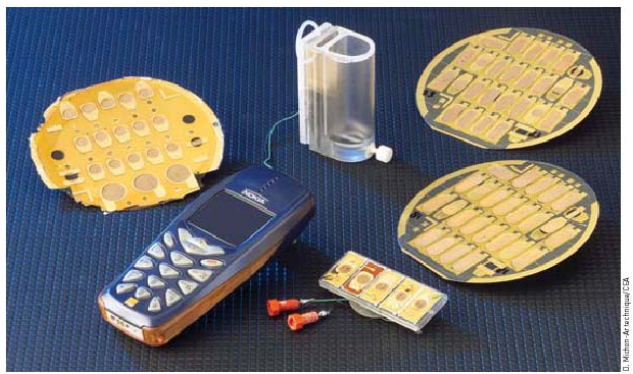

**Figure 1**: From the wafer to the phone, the fuel cells by the CEA *after [20]*.

Concerning near commercial applications, only announcements allow us to know a few details on FC about to come on the market. For instance, the French Nuclear Research Center CEA announced the successful fabrication of high performances prototypes (fig. 1) fuelled by $H_2$ and based on thin films type structures on Si substrate obtained by microelectronics fabrication techniques (RIE for fuel microchannels, PVD for anode collector, CVD, serigraphy, inkjet for Pt catalyst, lithography) with a Nafion® membrane [20]. An impressive power density of 300 mW cm$^{-2}$ with a stabilization around 150 mW cm$^{-2}$ during hundred hours was reported.

Since many years now, Morse *et al* [21] (Lawrence Livermore National Laboratory) have been working on Solid Oxide FC (SOFC) and PEMFC. They combined thin film and MEMS technologies to fabricate miniature FC. A silicon modular design served as the platform for FC based on either PEM or SO membrane. In 2002, the PEM cell yielded a computed peak power of 37 mW cm$^{-2}$ at 0.45 V and 40°C whereas the SOFC reached 145 mW cm$^{-2}$ at 0.35 V and 600°C. Since 2002, the UltraCell company has an exclusive license with Lawrence Livermore National Laboratory to exploit their technology to develop commercial FC.

## 2.2. The metal way

Less expensive than silicon and easier to assemble with other FC components, micromachined foil materials are also one possible choice to obtain low-cost functional FC.

Using the same structure as previously quoted [13,14], Lu *et al* have recently replaced the Si judged too fragile for compressing and good sealing with the MEA by very thin stainless steel plates as bipolar plates with the flow field machined by photochemical etching technology [22]. A gold layer was deposited on the stainless steel plates to prevent corrosion. This enhanced FC reached 34 mW cm$^{-2}$ at RT and 100 mW cm$^{-2}$ at 60°C.

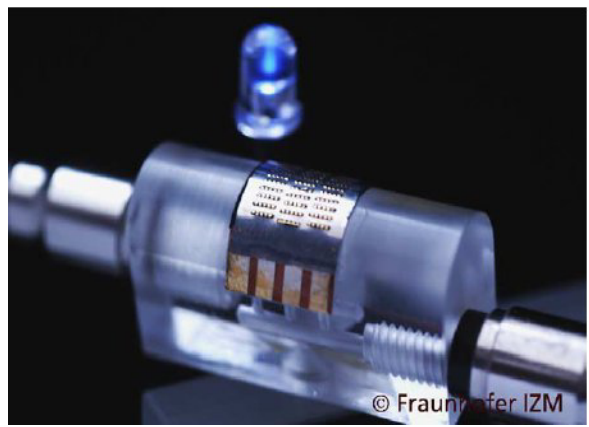

**Figure 2**: Micro fuel cell powering a LED, *after [23]*.

R. Hahn and the Fraünhofer IZM have also chosen foil materials for their prototypes of self-breathing planar PEMFC (one example in fig. 2) [23]. The dimensions are 1 cm² with 200 µm thickness. Stable long term operation was achieved at 80 mW cm$^{-2}$ at varying ambient conditions with dry $H_2$ fuel. They use a commercially available MEA further processed in their laboratory. Micro patterning technologies were employed on stainless steel foils for current collectors and flow fields and the assembled technology was adapted from microelectronics packaging. In their design, the interconnection between cells is performed outside the membrane area which reduces sealing problems. The complete FC consists of only 3 layers: the current collectors with integrated flow fields on top and bottom and the patterned MEA in-between. The current collector foils with integrated flow field of the electrodes consist of Au-electroplated and microstructured (micro patterning, wet etching, laser cutting, RIE) sandwiched metal-polymer foils. The commercial MEA is structured using RIE and laser ablation to avoid possible internal bypasses.

Müller *et al* (IMTEK) used micro-machined metal foils to form the flow fields of their µ-FC [24]. Using Gore MEA, they were able to form very thin, high power density stacks. They demonstrated both uncompressed and compressed FC designs. The uncompressed one had a peak power density of 20 mW cm$^{-2}$ and the 5-cell compressed stack achieved 250 mW cm$^{-2}$.

Though they do not refer to microfabrication techniques, we can also report another interesting metallic miniature FC structure once again coming from the University of Illinois and proposed by Ha *et al* [25]. They described the design and performance of a passive air breathing direct formic acid FC (DFAFC). The MEA was fabricated in house with catalyst inks directly painted on a Nafion®-117 membrane. The current collectors were fabricated from Ti foils electrochemically coated with gold. The miniature cell at 8.8 M formic acid produced a maximum power density of 33 mW cm$^{-2}$ with pieces of gold mesh inserted between the current collector and the MEA on both sides of the MEA. With Pd black used as the catalyst at the anode side [26], they recently obtained a maximum power density of 177 mW cm$^{-2}$ at 0.53 V for their passive DFAFC with 10 M formic acid.

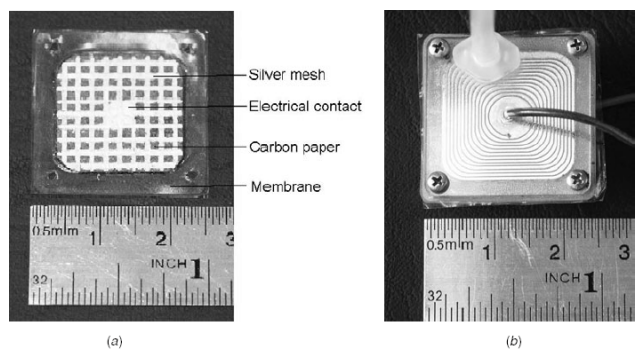

**Figure 3**: Assembled polymeric micro fuel cell, *after [27]*.
(a) open view and (b) assembled view.

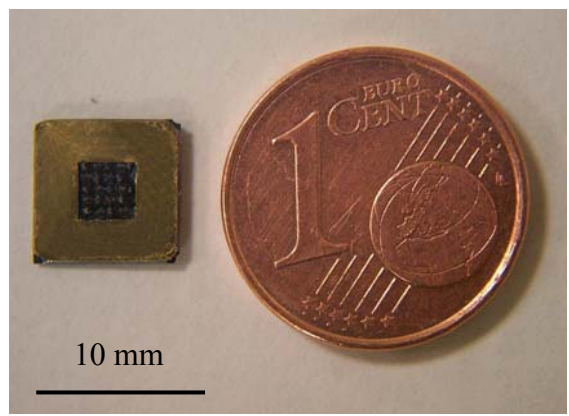

**Figure 4**: Membrane-electrode assembly (8×8 mm), scale comparison with a 1 cent euro coin.

**2.3. The polymer way**

A third solution currently developed speculates on MEMS-polymers such as polydimethylsiloxane (PDMS) or polymethyl methacrylate (PMMA). These polymers can be micromachined by molding or by laser ablation.

Chan *et al* (Nanyang Univ.) reported the fabrication of a polymeric µPEMFC (fig. 3) developed on the basis of micromachining of PMMA by laser [27]. The microchannels for fuel flow and oxidant were ablated with a $CO_2$-laser. The energy of the laser beam has a Gaussian distribution thus the cross section of the channel also has a Gaussian shape. A 40 nm gold layer was then sputtered over the substrate surface to act both as the current collector and corrosion protection layer. The Gaussian shape allows gold to cover all sides of the channel. In this µ-FC, water generated by the reaction was utilized for gas humidifying. The flow channel has a spiral shape which enables the dry gas in the outer spiral line to become hydrated by acquiring some of the moisture from the adjacent inner spiral line. The MEA consists in a Nafion® 1135 membrane with a hydrophobic carbon paper and a diffusion layer (carbon powder and PTFE) on one side and coated with a catalyst layer (ink with Nafion®-112 solution and Pt-carbon) on the other side. Silver conductive paint was printed on the other side of the carbon paper to increase its conductivity and to contribute to collect current. In the final step, the two PMMA substrates and the MEA were bonded together using an adhesive gasket. $H_2$ was supplied by hydride storage, the air by an air pump, all tests were performed at RT. A high power density of 315 mW cm$^{-2}$ at 0.35 V has been achieved.

Shah *et al* [28,29] have developed a complete PEMFC system consisting of a PDMS substrate with micro flow channels upon which the MEA was vertically stacked. PDMS microreactors were fabricated by employing micromolding with a dry-etched silicon master. The PDMS spin-coated on micromachined Si was then cured and peeled off from the master. The MEA employed consisted in a Nafion-112 membrane where they have sputtered Pt through a Mylar mask. Despite an interesting method, this FC gave poor results (0.8 mW cm$^{-2}$).

This brief overview shows that it is not the material that really matters to achieve high performances but mostly the way the fluids are managed and the stack is sealed. If Si is still the material the most commonly used, the promising performances and technology obtained by the Fraunhöfer Institute with stainless steel foils, for instance, have shown that interesting alternative technologies can also lead to functional FC.

We can also notice that besides the use of microfabrication techniques, all the different presented technologies have a major point in common which is the use of Nafion® or Nafion®-like ionomer as the proton exchange membrane.

We propose another approach consisting in using standard MEMS techniques and chemical grafting to create a new proton-conducting membrane.

## 3. GRAFTED POROUS SILICON-BASED MINIATURE FUEL CELLS

Our work focuses on the fabrication of new PEM for small FCs. The solution we developed as an alternative to classical ionomeric membranes like Nafion® is based on a PS membrane on which is grafted a proton-conducting silane.

Previous works have reported the relevance to use PS directly as the membrane of the FC, either as the support for a Nafion® solution [8] or with appropriate proton-conducting molecules grafted on the pore walls [30]. This second solution involves the chemical grafting of silane molecules containing sulfonate ($SO_3^-$) or carboxylate ($COO^-$) groups on the pores walls in order to mimic the structure of an ionomer.

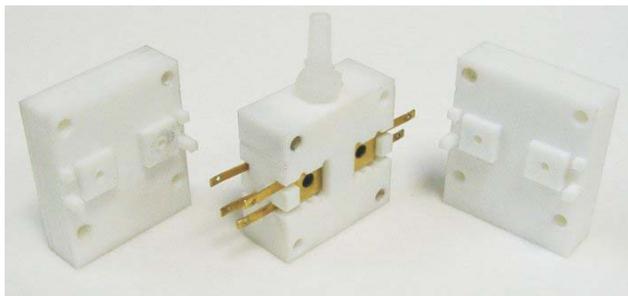

**Figure 5**: Home-made test system for 4 fuel cells.

The complete process for the fabrication of the proton-conducting membranes, previously reported in [30], can be summarized as follows. A $p^+$ (100)-oriented silicon wafer is first thermally oxidized, and covered with a sputtered Cr-Au layer on both sides. This is followed by classical photolithography (with adapted etch for the metal layers) and chemical wet etching by KOH to produce 50 µm thick double-sided membranes. These membranes are then made porous by anodization in an ethanoic HF bath with an average pore diameter of 10 nm.

PS is not naturally proton-conducting. In order to make the membranes proton-conducting, our solution consists in grafting silane molecules containing proton-conducting groups on the pores walls. We have used a commercially available silane salt from United Chemical Technologies Inc (UCT) containing three carboxylate groups for the first investigations. The first step consists in creating silanol functions (Si-OH) at the surface of PS. A soft process involving UV ozone cleaner has been successfully implemented. The grafting of silane molecules is then realized by immersing the hydrophilic porous membranes into a 1 % silane solution in ethanol for 1 h at RT and ambient air. In order to replace -Na endings by -H endings to get the real carboxylic behaviour for the grafted function, membranes are immersed for 12 hours in a 20 % solution of sulfuric acid, then carefully rinsed in deionized water. To complete the FC assembly, E-tek carbon conducting cloth electrodes filled with Pt (20 % Pt on Vulcan XC-72) are added on both sides of the membrane as a $H_2 / O_2$ catalyst.

### 4. RESULTS

Figure 4 shows (top view) a typical 8 mm × 8 mm FC realized with an active area (in black on the figure) of 7 mm².

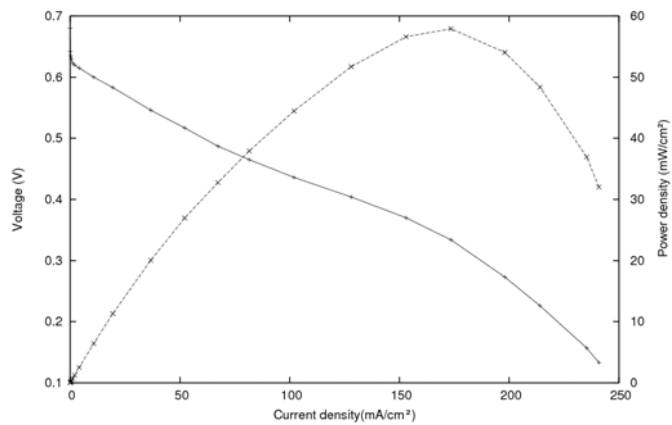

**Figure 6**: Performances obtained with a mesoporous silicon membrane with grafted silane (voltage vs current density in normal line, power density vs current density in dashed line).

Measurements were carried out at RT. $H_2$ feeding is provided by a 20 % NaOH solution electrolysis while $O_2$ is provided directly by passive ambient air. In order to bring the gas to the membrane, a new home-made test cell was used (fig. 5) in which up to 4 MEA can be tested separately or together with a serial or parallel connexion. With an improved MEA and a better test-cell but with the same grafted silane, performances have been considerably increased compared to previously reported works [30]. A maximum power density of 58 mW cm$^{-2}$ at 0.34 V has been reached with an open circuit voltage (OCV) of 680 mV (fig. 6) when only 17 mW cm$^{-2}$ and an OCV of 480 mV had been previously measured. This can be explained by a better grafting technique which enables to reduce the crossover of $H_2$ by shrinking the open pores.

In order to further reduce this crossover and then to increase the OCV, the pore diameter should be reduced as small as that of assumed pores in Nafion®. Moreover, to further increase the power density, the grafting density should be controlled to be sure that all the internal surface of the pores is grafted.

### 5. CONCLUSION

We have presented a few examples of miniature FC using microfabrication techniques. Among the numerous solutions developed today, the basic structure of FC remains the same: thin film planar stack (silicon, foils, polymer or glass) with commercial ionomer, most often Nafion®, the reported layers being micromachined (microchannels or porous media) for gas/liquid management and coated with gold for current collecting. Performances range from the tenth of µW cm$^{-2}$ to several hundreds of mW cm$^{-2}$ according principally to the fluids management and the sealing. The Si remains the most employed material, but foils and polymers have shown interesting perspectives for future commercial developments. We have also reported an alternative solution which does not use an ionomer for the PEM but consists in a PS membrane with a proton-conducting silane grafted on the pores walls. With this membrane, performances as high as 58 mW cm$^{-2}$ have been achieved. This promising membrane can still be improved. Future work should focus on the reduction of the pores diameter to decrease the gas crossover, the control of the grafting density into the porous silicon, the replacement of the electrode carbon cloth by an ink and the use of a more proton-conducting silane (with $SO_3^-$ terminations).